\documentclass {jfm}
\usepackage{graphicx}
\usepackage{natbib,amssymb,amsmath}
\usepackage{color}
\usepackage{enumitem}
\setlist[description]{leftmargin=1.4\parindent,labelindent=0pt}

\newcommand{\red}{\textcolor{black}}
\newcommand{\rev}{\textcolor{black}}

\def\e{\mathrm{e}}
\def\Re{\operatorname{Re}}
\def\Im{\operatorname{Im}}

\renewcommand{\theequation}{\thesection.\arabic{equation}}
\let\ssection=\section
\renewcommand{\section}{\setcounter{equation}{0}\ssection}

\def\Re{\mathrm{Re}\,}

\title[Inertia-gravity waves in shear  flows]{
Inertia-gravity waves in inertially stable and unstable  \red{shear} flows}
\author[F. Lott, C. Millet and J. Vanneste]
{Fran\c{c}ois Lott$^1$\thanks{Email address for correspondence: flott@lmd.ens.fr},
Christophe Millet$^2$,
and Jacques Vanneste$^3$}
 
\affiliation{$^1$Laboratoire de M\'et\'eorologie Dynamique du CNRS,
Ecole Normale Sup\'erieure,
24, rue Lhomond, 75231 Paris cedex 05, France\\[\affilskip]
\red{$^2$ CEA, DAM, DIF, 91297 Arpajon, France}\\[\affilskip]
$^3$School of Mathematics and Maxwell
Institute for Mathematical Sciences,    
University of Edinburgh, Edinburgh EH9 3FD, UK}

\date{November 20, 2014}

\begin{document}

\maketitle




\begin{abstract}
An inertia-gravity wave (IGW) propagating in a vertically sheared, rotating stratified fluid interacts with the pair of inertial levels that surround the critical level. An exact expression for the form of the IGW is derived here in the case of a linear shear and used to examine this interaction in detail. This expression recovers  
the classical values of the transmission and reflection coefficients 
$|T|=\e^{-\pi \mu }$ and $|R|=0$, where $\mu^2=J(1+\nu^2)-1/4$,
 $J$ is the Richardson number
and $\nu$  the ratio between the horizontal transverse and along-shear
wavenumbers.

For large $J$, a WKB analysis provides an interpretation of this result
in term of tunnelling: an IGW incident to the lower inertial level becomes evanescent between the inertial levels, returning to an oscillatory behaviour above the upper inertial level. The amplitude of the transmitted wave is directly related to the decay of the evanescent solution between the inertial levels. In the immediate vicinity of the critical level, the evanescent IGW is well represented by the quasi-geostrophic approximation, so that the process can be interpreted as resulting from the coupling between balanced and unbalanced motion.


The exact and WKB solutions describe the so-called valve effect, a dependence of the behaviour in the region between the inertial levels on the direction of wave propagation. For $J < 1$ this is shown to lead to an
amplification 
of the wave between the  inertial levels. Since the flow is inertially unstable for $J<1$, this establishes a correspondence between the inertial-level interaction  and the condition for inertial instability.

\end{abstract}

\section{Introduction}

Inertia-gravity waves (IGWs), that is, internal gravity waves with frequencies close enough to the Coriolis frequency to be affected by rotation, are ubiquitous
in both the ocean and the atmosphere (for recent observations
see Marshall et al.~2009 and Hertzog et al.~2008, respectively).
In the ocean, they are efficiently excited by 
surface winds especially in the region
of the atmospheric storm tracks (Alford 2003). In the atmosphere,
external sources such as mountains and convection
produce IGWs (Scavuzzo et al.~1998, Lott~2003);
internal sources, like those appearing
during the life cycle of baroclinic instabilities are also efficient 
(Plougonven and Snyder~2007, Sato and Yoshiki 2008).
The impact of atmospheric IGWs on the global climate
is now well established 
\red{(Andrews et al.~1987)}, and a current challenge is the quantification of the non-convective
sources that are parameterized in middle atmosphere models
(Zuelicke and Peters 2008, Richter et al.~2010). 
In the ocean, IGWs are important for several processes including small-scale mixing and energy dissipation (Ferrari and Wunsch 2009).

IGWs are generated in or propagate through regions where intense jets exist (Whitt and Thomas 2013), and they interact with the jets.
These interactions involve critical levels, that is, regions where the Doppler-shifted 
frequency of the wave is zero (Booker and Bretherton~1967)
or, when rotation is taken into account, inertial levels where the Doppler-shifted frequency 
equals the Coriolis frequency
(Jones 1967). Such interactions arise, for instance,
when a low-level front passes over a mountain ridge, yielding directional
critical levels almost everywhere at low altitudes (Shutts 2003, Shen and Lin 1999).

A central result, valid both with and without rotation, is that an IGW originating from $z \to -\infty$, propagating though critical and inertial levels and radiating away as $z \to \infty$, is not reflected and is absorbed by a factor
\begin{equation} 
|T|=\exp(-\pi \mu),  \quad \textrm{where} \ \ \mu = \sqrt{J(1+\nu^2)-1/4}, \label{eq:BB_Abs}
\end{equation}
$J > 1/4$ is the Richardson number, and $\nu=l/k$ is the ratio of the horizontal components of the wavevector respectively transverse to and aligned with the shear. 
This result is obtained for an inviscid, hydrostatic fluid under the assumption of uniform vertical shear and Brunt--V\"ais\"al\"a frequency. It gives the ratio of the amplitude of the transmitted wave to the incident wave  and, in the rotating case, conceals a complex  behaviour between the inertial levels. This is exemplified by the fact that the change of the wave amplitude across the lower and upper inertial level are given by factors
\begin{equation}
\exp\left( \pi \nu  \right) \quad \textrm{and} \quad
\exp\left( - \pi \nu  \right),
 \label{eq:JO_Abs}
\end{equation}
respectively (see Grimshaw 1975),
and by the fact that the wave amplitude does not change at the critical level.
While they cancel overall, the factors in (\ref{eq:JO_Abs}) reveal a strong dependence of the wave amplitude between the two inertial levels on the direction $\nu=l/k$ of the wave vector. This phenomenon, referred to as the ``valve effect'', has been considered in earlier papers  (Grimshaw~1975, Yamanaka and Tanaka~1984), but none are entirely satisfactory
since they do not provide exact solutions or approximate solutions
over the entire vertical axis. In this respect it is worth recalling that the overall transmission in (\ref{eq:BB_Abs})
 is the same with or without rotation (Jones~1967) because the solutions as $z \to \pm \infty$ can be connected  by integrating the relevant differential equation  along a path with $|z| \gg 1$ on which the  solution is asymptotically unaffected by rotation (see Fig.~\ref{fig:TP}).  
This is made possible by the structure of the Stokes lines which, unlike in the familiar Airy equation for instance, do not go to $\infty$, \red{but simply join the two inertial levels along the real $z$-axis}. 
 One of our aims is to give an exact solution for $z=O(1)$ that displays a dependence on rotation and $\nu$; another is to provide a description of the solution that reconcile the absorptive properties in (\ref{eq:BB_Abs}) and (\ref{eq:JO_Abs}).

Another motivation for the paper is the relation between critical and inertial levels, and instability. This relation is well known: classical shear instabilities, for instance, can produce internal gravity waves (Lott et al. 1992, Lott~1997), in which case a critical level is necessarily present (Miles 1961); conversely, the interaction at a distance between gravity waves helps explain stratified shear flow instability (Rabinovitch et al. 2011). 
Similarly, recent works on the coupling between balanced waves and unbalanced waves show that critical and inertial levels are associated with unbalanced instabilities (Sakai 1989, Molemaker et al. 2005, Plougonven et al. 2005, Vanneste and Yavneh 2007, Sutyrin 2008, Gula and Zeitlin 2010). In the ocean, these unbalanced instabilities may have an important role in providing dynamical routes to dissipation at small scales (McWilliams 2003).

In this context of flow stability, the fact that
(\ref{eq:BB_Abs}) remains valid in the rotating case is surprising.
Indeed, in the non-rotating case the conditions for gravity-wave absorption
and flow stability are the same,  $J>1/4$
(Miles 1961, Howard 1961), leading to an interpretation of instability in terms of gravity-wave amplification  (Lindzen 1988). In the presence of rotation, however, symmetric instability (with $k=0$) occurs for $J<1$ (Stone 1966, Bennetts and Hoskins 1979), \red{but the
critical value
$J=1$ does} not directly relate to the wave-absorption properties in (\ref{eq:BB_Abs})
or (\ref{eq:JO_Abs}).
For $J \gtrsim 1$ and with horizontal boundaries, non-symmetric disturbances can compete
with the symmetric ones (Mamatsashvili et al.~2010), but again
the relation with the absorptive properties of the critical layer is not clear. 
One hint of a relation for symmetric and near-symmetric disturbances is the behaviour of the amplification factor in (\ref{eq:JO_Abs}) as $\nu = l/k \rightarrow\infty$, but this does not involve $J$.

The purpose of this paper is to reconcile
the absorptive properties in (\ref{eq:BB_Abs}) and (\ref{eq:JO_Abs})
by giving an exact solution 
of the Taylor--Goldstein equation, with constant shear,
stratification and without boundaries, that is valid over the entire range of altitude
in the rotating case.
The interpretation of this solution is facilitated by an asymptotic WKB treatment valid when $J \gg 1$. 
Our method is related to that of Lott et al~(2010, 2012 hereinafter LPV12) who showed the importance of
inertial levels for the spontaneous emission of gravity waves
by potential-vorticity (PV) anomalies. Specifically, they showed that a PV anomaly in a shear 
induces a balanced response that decays exponentially in the vertical up to the inertial levels, then takes the form of a propagating IGW. 

In the present paper, we consider an IGW that is incident from below to a pair of inertial levels.
It changes nature across the lower inertial level, taking an exponentially decreasing, balanced form, and is restored to an IGW of much weaker amplitude across the upper inertial level. 
Qualitatively, the amplitude of
the transmission coefficient in (\ref{eq:BB_Abs}) is set by the value of the
balanced solution immediately below the upper inertial level.
Our exact and approximate results provide the properties of the solution for a broad range of values of $J$ and give a complete description of the solution between the inertial levels, where the valve effect is manifest.
For  $J<1$, in particular, we identify a form of disturbance amplification, whereby the Eliassen--Palm (EP) flux
of the wave is amplified as the wave crosses the lower inertial level. 
This establishes a clear correspondence between the absorptive properties
of the shear layer and the criterion $J<1$ for flow instability.

The plan of the paper is as follows.
The exact solution for the vertical structure of the wave is derived in section 2. As the mathematics are quite involved
but present only few conceptual difficulties much of the derivation
is relegated to an Appendix.
Section 3 develops an interpretation of the exact result in terms of tunnelling using
a WKB analysis for $J \gg 1$. The inertially unstable case $J<1$ is analysed in section 4. The paper concludes with a brief discussion in  section 5.

\section{Exact solution}

We start from the linearized hydrostatic--Boussinesq equations for the propagation of
a  three-dimensional disturbance in a uniformly stratified sheared
flow $\mathbf{\overline{u}_0}=(\Lambda z, 0, 0)$,
where the shear $\Lambda>0$ and the Brunt-V\"ais\"al\"a frequency $N$
are constant. When the Coriolis parameter $f$ is also constant, \red{a stationary monochromatic disturbance with  vertical velocity of the form}
\begin{equation}
w(x,y,z,t)=\Re\left\{W(z)\e^{i\left(kx+ly-\omega t\right)}\right\}, 
\label{eq:wmono}
\end{equation}
with $\omega$ the absolute frequency,
satisfies the vertical structure equation
\begin{equation}
\frac{1-\xi^2}{\xi^2} W_{\xi\xi}
-\left(\frac{2}{\xi^3}-\frac{2i\nu}{\xi^2}\right)W_\xi
-\left[\frac{(1+\nu^2)J}{\xi^2}+\frac{2i\nu}{\xi^3}\right]W=0,
\label{eq:deltay}
\end{equation}
where
\begin{equation}
J=\frac{N^2}{\Lambda^2}, \; \; 
\nu=\frac{l}{k} \; \; \ \mbox{and} \; \; \
\xi=\frac{k\Lambda z}{f}-\frac{\omega}{f}. \label{eq:xi}
\end{equation}
\red{This is  Eq.~(6) in Yamaka and Tanaka~(1984), but see also Inverarity and Shutts~(2000).
Note that in the absence of horizontal boundaries, as assumed here, and for $k\not=0$, there are no unstable modes so we can take $\Im  \omega =0$; for $k=0$, inertial instability occurs for $J<1$, leading to growing modes with $\Im \omega \not= 0$ which we discuss further in section 4.}

Equation (\ref{eq:deltay}) has two singularities \red{at the inertial levels} $\xi=\pm 1$ that
require a special treatment; \red{the critical level} $\xi=0$ is an apparent
singularity. Following LPV12, we  write
the exact solutions of (\ref{eq:deltay}) as
\begin{subequations}
\begin{equation}
W^{(\mathrm{III})}_{u}(\xi)\; \; \mbox{for} \; \xi>1, 
\label{eq:sol1}
\end{equation}
\begin{equation}
A W^{(\mathrm{II})}_d(\xi)+B W^{(\mathrm{II})}_u(\xi)  \; \; \mbox{for} \; -1<\xi<1,
\label{eq:sol2}
\end{equation}
\begin{equation}
C W^{(\mathrm{I})}_d(\xi)+D W^{(\mathrm{I})}_u(\xi)  \; \; \mbox{for} \; \xi< -1.
\label{eq:sol3}
\end{equation}
\end{subequations}
In (\ref{eq:sol1})--(\ref{eq:sol3}) $A$, $B$, $C$ and $D$ are constants,
and the $W$ functions can be expressed in terms of hypergeometric functions
(see Appendix,  Yamanaka and Tanaka 1984 or 
Shutts~2001).
In (\ref{eq:sol1}) we have retained the solution with the
asymptotic form 
\begin{equation}
\label{eq:uprad}
W^{(\mathrm{III})}_u(\xi) \sim \xi^{1/2 +i\mu} \quad \textrm{as} \  \, \xi \to \infty,
\end{equation}
corresponding to an upward-propagating gravity wave
(Booker and Bretherton, 1967).
In (\ref{eq:sol3}) the two functions are such that 
the asymptotic form is given by
\begin{equation}
W^{(\mathrm{I})} \sim C|\xi|^{1/2-i\mu}+D|\xi|^{1/2+i\mu} \quad \textrm{as} \  \, \xi \to -\infty.
\label{eq:OPV_asyb}
\end{equation}

\begin{figure}
\begin{center}
\includegraphics[width=14cm]{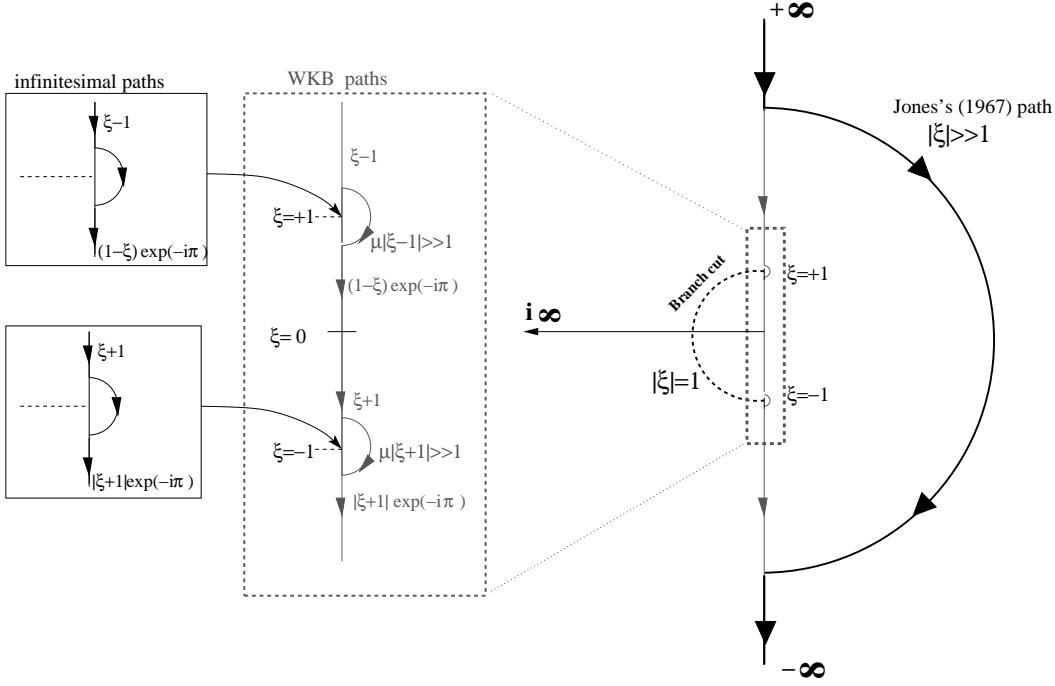}
\end{center}
\caption{Paths  \red{used for the integration of} 
the vertical structure equation (\ref{eq:deltay}); \red{see also Figure~1 in Jones~(1967) and note that our 
real $\xi$-axis is vertical while Jones's is horizontal.} }
\label{fig:TP}
\end{figure}

To evaluate $A$, $B$, $C$, and $D$, we first use asymptotic expansions of (\ref{eq:sol1})--(\ref{eq:sol2}) and (\ref{eq:sol2})--(\ref{eq:sol3})
in the vicinity of $\xi=1$ 
and $\xi=-1$, respectively (see Appendix). Next we follow Booker and Bretherton (1967) and introduce an infinitesimaly small linear damping to determine the physically relevant branches of multivalued functions (see Fig.~\ref{fig:TP}).
This is equivalent to shifting the \red{path of integration of (\ref{eq:deltay})}
into the lower half of the complex plane so that
\begin{equation}
\begin{array}{rcl}
\xi-1=(1-\xi) \e^{-i\pi} & \quad \textrm{for} \quad & \xi < 1, \quad \mbox{and} \\
\xi+1=|\xi +1| \e^{-i\pi} &  \quad  \mbox{for} \quad & \xi<-1.
\end{array}
\label{eq:branchs}
\end{equation}
Using this, the asymptotic expansions of (\ref{eq:sol1}) and (\ref{eq:sol2}) can be continued for $\xi \to 1^-$ and
$\xi \to -1^-$ and compared with the corresponding  expansions of (\ref{eq:sol2}) and (\ref{eq:sol3}) \red{(see the Appendix for an illustration of the procedure in the simpler context of the WKB solution)}. 
Matching expansions near $\xi=1$ gives 
\begin{equation}
\begin{array}{rcl}
\alpha' & = & \e^{-\nu\pi} \left(\alpha A+\alpha'' B \right),\\
\beta' &= &\beta  A+\beta'' B,
\end{array}
\label{eq:system1}
\end{equation}
where the coefficients $\alpha, \, \beta, \alpha',$etc are known functions of $\mu$ and $\nu$ given in the Appendix.
Similarly, matching near $\xi=-1$ gives
\begin{equation}
\begin{array}{rcl}
\beta A -\beta''B & = &{\e}^{\nu \pi} \left( \beta''' C+\beta'D \right),\\
\alpha'''C+\alpha'D & = &\alpha A -\alpha ''B.
\end{array}
\label{eq:system2}
\end{equation}
The constants $A, \, B, \, C$ and $D$ are then deduced by solving the linear system (\ref{eq:system1})--(\ref{eq:system2}). Note that the valve effect, involving an exponential dependence on $\nu$, can be traced to the factors  $\e^{\pm \nu\pi}$ in (\ref{eq:system1})--(\ref{eq:system2}), with the amplification across the lower inertial level associated with the factor in (\ref{eq:system2}) and the balancing attenuation across the upper critical level associated with that in (\ref{eq:system1}).

\begin{figure}
\begin{center}
\includegraphics[width=8cm]{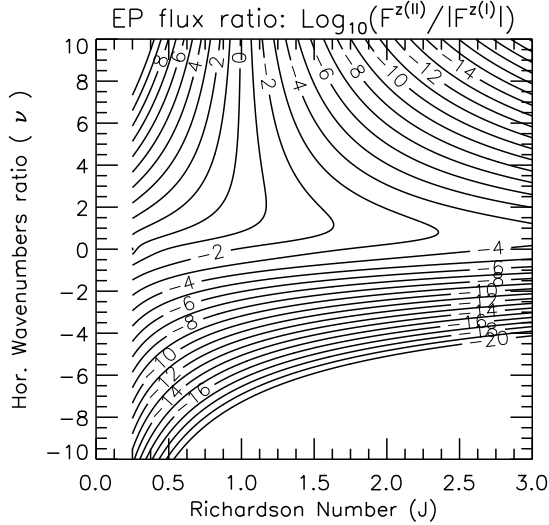}
\end{center}
\caption{\red{Ratio of the EP flux  between the inertial levels
to the incident EP flux as a function of the Richardson number $J$ and wave aspect ratio $\nu = l/k$}.}
\label{fig:EP_FLUX}
\end{figure}

After involved manipulations it can be shown that the reflection and transmission
coefficients are given by
\begin{equation}
R=\frac{C}{D}=0\quad \hbox{and}\quad T=\frac{1}{D}=i\e^{-\mu \pi},
\label{eq:RandT}
\end{equation}
where $\mu^2=J(1+\nu^2)-1/4$ (see Appendix A.1).
This is the result of Jones~(1967), which can be obtained much more directly by integration along a semi-circle with $|\xi| \gg 1$, as recalled in the introduction (see Fig.~\ref{fig:TP}). Our solution provides the details of the solution between the inertial levels that this result ignores. In particular, it \red{makes it possible to compute the EP flux, proportional to} 
\begin{equation}
F^z=
 \mathrm{Re} \, \left( i \frac{1-\xi^2}{\xi^2} W_\xi W^*-\nu \frac{W W^*}{\xi^2} \right),
\label{eq:EP_flux}
\end{equation}
and constant apart from jumps at the inertial levels \red{(this conservation law is derived directly from (\ref{eq:deltay})
by multiplying by $W^*$ and integrating by parts).
We can obtain the three distinct values of the EP flux on either side of and between the inertial levels using the first terms in the expansion of the hypergeometric functions for $\xi\to\pm\infty$ and 
$\xi\to 0$. Computations detailed in the Appendix show that} the ratios of the EP flux between and above the inertial levels to the EP flux below are
\red{
\begin{equation}
\frac{F^{z(\mathrm{II})}}{|F^{z(\mathrm{I})}|} = \frac{\e^{2\pi\nu}+1}{\e^{2\pi\mu}-1} \quad \textrm{and} \quad
\frac{F^{z(\mathrm{III})}}{|F^{z(\mathrm{I})}|} = \e^{-2\pi\mu}, \label{eq:epratio}
\end{equation}
where the absolute value in the denominators is required because the incoming 
EP flux $ F^{z(\mathrm{I})}<0$ whereas the other two fluxes are positive (see Appendix).}
The first of these ratios describes the crossing of the lower inertial level by the incident waves and displays the strong dependence on the direction $\nu$ of the wavevector that characterises the valve effect. It is shown as a function of $\nu$ and $J$ in Fig.~\ref{fig:EP_FLUX}. The figure indicates that, for $J<1$, the ratio can exceed 1 and can take very large values for $\nu > 0$. We discuss this phenomenon further in section 4.

\section{WKB approximation and tunnelling}

A WKB solution of (\ref{eq:deltay}), strictly valid for $\mu \gg 1$ but qualitatively correct for $\mu \gtrsim 1$, sheds light on the behaviour of the solution and on the valve effect in particular. To derive this solution, it is convenient to follow  Miyahara~(1981) and recast (\ref{eq:deltay}) in the canonical form
\begin{subequations}
\begin{equation}
\Psi_{\xi\xi}+\left[\frac{\mu^2+{1}/{4}}{\xi^2-1}
+\frac{(3+\nu^2)\xi^2-2}{\xi^2(\xi^2-1)^2}\right]\Psi=0,
\label{eq:canon}
\end{equation}
where
\begin{equation}
W(\xi)=\xi(\xi-1)^{-1/2+i\nu/2}\left(\xi+1\right)^{-1/2-i\nu/2} \Psi(\xi).
\label{eq:psidef}
\end{equation}
\end{subequations}
Assuming  $\mu \gg 1$, we then introduce the  expansion
\begin{equation}
\Psi(\xi)=\left(\Psi_0(\xi)+\mu^{-1}\Psi_1(\xi)+...\right)
\e^{\mu\int^\xi \phi(\xi')d\xi'}
\label{eq:WKB_exp}
\end{equation}
\red{and find at orders $\mu^2$, $\mu$ and $1$} that
\begin{equation}
\phi=\frac{ \epsilon i}{\sqrt{\xi^2-1}}, \ \  \Psi_0=\left(\xi^2-1\right)^{1/4} \ \ \textrm{and} \  \ \Psi_1=-\epsilon i \frac{1+(\nu^2/2-7/8)\xi^2}{\xi(\xi^2-1)^{1/4}},
\label{eq:order1}
\end{equation}
where $\epsilon=\pm 1$ selects the two possible solutions.
The expansion (\ref{eq:WKB_exp}) \rev{breaks down  in the regions 
$||\xi|-1| = O(\mu^{-2})$} surrounding the inertial levels where an approximation in terms of Hankel functions can be constructed (see LPV12 \red{for the rescaling leading to this conclusion}).
\red{Here we avoid this complication by integrating (\ref{eq:canon})
 along a path
that remains at a distance larger than $\mu^{-1}$ from
the inertial levels 
and avoids the branch cut
(see Fig.~\ref{fig:TP}).}
\red{This path does not cross the only Stokes line (where $\Im \phi = 0$, e.g., Ablowitz \&  Fokas 1997), namely $\xi \in [-1,1]$. Therefore,  
%
a single WKB solution, determined by the radiation condition to correspond to $\epsilon = 1$, can be continued from $\xi \to \infty$ to $\xi \to - \infty$.}
To make the absorptive properties more transparent, we express the approximations at  order $\mu$
for a unit-amplitude incident wave
 (rather than for a unit-amplitude \red{transmitted wave} in
the exact case, see Appendix A.3 for details). \red{This gives}
\begin{subequations} \label{eq:WKB}
\begin{equation}
\begin{array}{lcrccr}
W_{\mathrm{WKB}}^{(\mathrm{III})}= & \hspace{8mm}  &-  i g(\xi) & \e^{-\mu\pi} &
\e^{i\mu\ln\left(\xi+\sqrt{\xi^2-1}\right)} & \,\mbox{for} \; \xi>1,
\end{array}
\label{eq:WKB_III}
\end{equation}
\begin{equation}
\begin{array}{lcrccr}
W_{\mathrm{WKB}}^{(\mathrm{II})}= &- \e^{\nu\pi/2} &
\frac{1-i}{\sqrt{2}}
g(\xi) &
\hspace{6mm} & \e^{-\mu (\pi/2+\sin^{-1} \xi)} &
 \; \mbox{for} \;  -1<\xi<1, 
\end{array}
\label{eq:WKB_II} 
\end{equation}
\begin{equation}
\begin{array}{lcrccr}
W_{\mathrm{WKB}}^{(\mathrm{I})}= &  \hspace{14mm} &
 g(\xi) & \hspace{6mm}
& \e^{i\mu\ln\left(|\xi|+\sqrt{\xi^2-1}\right)} & \mbox{for}
\; \xi<-1.
\end{array}
\label{eq:WKB_I}
\end{equation}
\end{subequations}
\red{Here, the function
$g(\xi)=\xi|\xi-1|^{-\frac{1}{4}+i\frac{\nu}{2}}|\xi+1|^{-\frac{1}{4}-i\frac{\nu}{2}}$ 
groups the factors unaffected by the absorptive properties of the shear layer;
 because it is expressed in terms of absolute values of $\xi \pm 1$, the complex powers \rev{$-1/4\pm i \nu/2$} apply to purely positive real arguments, and the effect of the continuations
(\ref{eq:branchs}) appears explicitly through exponential factors in (\ref{eq:WKB_III})--(\ref{eq:WKB_II}).
Note that for $|\xi| \gg 1$, 
$\e^{i\mu\ln\left(|\xi|+\sqrt{\xi^2-1}\right)}\approx 2^{i\mu}|\xi|^{i\mu}$
and $ g(\xi)\approx \mbox{sign}(\xi)|\xi|^{1/2}$,  so $W_{\mathrm{WKB}}^{(\mathrm{I})} = |\xi|^{1/2+i\mu}$, recovering the incident wave in (\ref{eq:OPV_asyb}) (up to an irrelevant phase factor).}


The WKB solution (\ref{eq:WKB}) is instructive in several respects. 
First, as $\xi \to \pm \infty$, it reproduces exactly the ratio between the amplitudes of the upward-propagating IGWs  (\ref{eq:uprad}) and (\ref{eq:OPV_asyb})  found from the exact solution. In particular, the reflection and transmission coefficients are the exact ones in (\ref{eq:RandT}). The WKB solution also shows that the \red{amplitude of the solution is dominated by 
the exponential term $\e^{-\mu\sin^{-1}\xi}$ in  (\ref{eq:WKB_II}) that rapidly} decays monotonically between the inertial levels. 
This decay is strongly reminiscent of the exponential decay that characterizes 
the neutral solution in the quasi-geostrophic approximation.
This can be made more transparent
by noting that near the critical level $\xi=0$, the $O(\mu^{-1})$ term
(retaining $\Psi_1$ in (\ref{eq:WKB_exp}))
is proportional to the quasi-geostrophic solution
$\left(1 + \mu \xi \right) \e^{-\mu \xi}$
(see Eq.~(2.16) in LPV12). Away from the immediate vicinity of $\xi=0$, the decay is in fact faster than predicted by the quasi-geostrophic approximation since $| \sin^{-1} \xi | \ge \xi$.
The valve effect is also evident in the WKB solution, through the factor $\hbox{e}^{\nu\pi/2}$ in (\ref{eq:WKB_II}).  
The dependence of the amplitude on $\nu$ associated to this factor is however weak compared to the dependence on $J$ and $|\nu|$ stemming from the factor $\e^{-\mu (\pi/2 + \sin^{-1} \xi)} \sim \e^{-\sqrt{J(1+\nu^2)}  (\pi/2 + \sin^{-1} \xi)}$ in (\ref{eq:WKB_II}) since $J \gg 1$. This factor represents a very rapid amplitude decay of the solution above $\xi=-1$, \red{and corresponds to an absorption that}
increases rapidly with $J$ and $|\nu|$: the asymmetry that favours perturbations with $\nu>0$ over those with $\nu < 0$ between the inertial levels is of little significance compared to this absorption.

%
%
%

\begin{figure}
\begin{center}
\includegraphics[width=10cm]{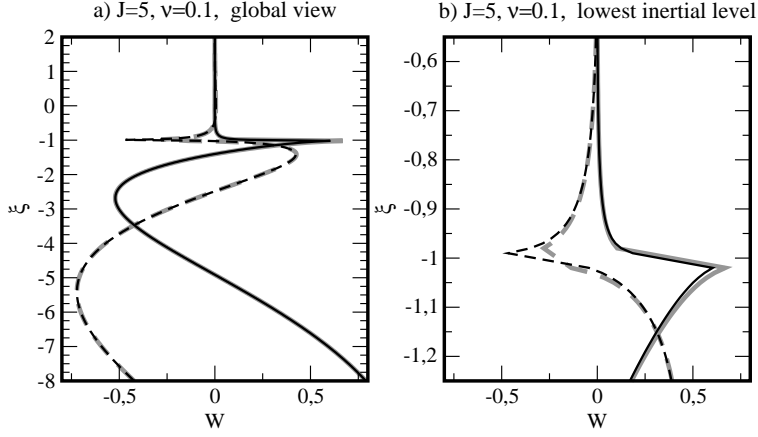}
\end{center}
\caption{Vertical velocity $W$ for the
exact solution~(\ref{eq:sol1})-(\ref{eq:sol3}) ($\Re W$, black solid; $\Im W$, black dashed) and the WKB approximation~(\ref{eq:WKB_III})-(\ref{eq:WKB_I})
($\Re W_{\mbox{\tiny WKB}}$ grey solid and $\Im W_{\mbox{\tiny WKB}}$ grey dashed) for $J=5$, $\nu=0.1$:
(a) global view; (b) zoom on the lowest inertial level.}
\label{fig:exact_vs_WKB} 
\end{figure}

The WKB solution represents
the exact solution accurately even for moderately large $J$. For $J = 5$, for instance, the two solutions 
almost coincide everywhere except in regions close to the inertial levels (see  Fig.~\ref{fig:exact_vs_WKB}).
The WKB approximation deteriorates around the lower inertial level for
$|\xi+1| \lesssim  0.05$ (see  Fig.~\ref{fig:exact_vs_WKB}b), 
corresponding to an $O(\mu^{-1})$ distance to the inertial level, as expected.
An analogous breakdown of the WKB approximation occurs for $\xi \approx 1$, but in this case
the solutions are very small. 

Figure~\ref{fig:exact_vs_WKB} also illustrates how the solution behaves as an upward-propagating wave -- with real and imaginary parts in quadrature -- outside the inertial levels (see Fig.~\ref{fig:exact_vs_WKB}a), and as an evanescent perturbation -- with real and imaginary parts in phase -- between the inertial levels (see Fig.~\ref{fig:exact_vs_WKB}b). This behaviour can be interpreted as a form of tunnelling, analogous to, but different from the classical tunnelling between turning points as described, for instance, by Bender and Orszag (1978). The continuity of the arguments of the last exponential factors in (\ref{eq:WKB}) shows that the  amplitude decrease of the transmitted signal by the factor  $\hbox{e}^{-\mu \pi}$ can be immediately related to the decay of the solution $W^{(\mathrm{II})}_{\tiny \mbox{WKB}}$ between the inertial level as $\e^{-\mu (\pi/2 + \sin^{-1} \xi)}$ since this decreases by the same factor $\hbox{e}^{-\mu \pi}$. 
In the tunnelling interpretation, the valve effect seems to play no role, 
because the amplitude jump $\hbox{e}^{\nu\pi/2}$ through the lower inertial level is exactly 
balanced by the inverse jump $\hbox{e}^{-\nu\pi/2}$ at the upper inertial level.
 
An obvious difference with classical tunnelling is the absence of a fully conserved flux. In classical tunnelling, wave-action flux conservation leads to the relation $|T|^2 + |R^2| =1$ that constrains the transmission and refection coefficients. Here, the analogous conservation is that of the EP flux, but because this can jump at the inertial level (Eliassen and Palm~1961, Grimshaw~1975), there is no associated constraint on $T$ and $R$, making the absence of reflection $R=0$ compatible with partial (indeed very weak) transmission $|T| < 1$. Regarding the EP flux, we note that the WKB solution between the inertial levels (\ref{eq:WKB_II}) suggests a zero flux, since it is evanescent (see for instance Gill~1982). The flux is non zero, however, and can be captured by extending the WKB analysis to take into account the Stokes phenomenon that arises at $\xi=1$. The relevant computation, carried out in LPV12, reveals the presence of an exponentially small solution increasing exponentially with $\xi$ to be added to (\ref{eq:WKB_II}). Taking this into account gives a non-zero flux, deduced from Eq.\ (2.31) in LPV12 to satisfy
\begin{equation}
\frac{F^{z(\mathrm{II})}}{|F^{z(\mathrm{I})}|} \sim \e^{-2 \pi \mu}(\e^{2\pi\nu}+1), \label{eq:epratiowkb}
\end{equation}
consistent with (\ref{eq:epratio}) for $\mu \gg 1$.

\section{Valve effect and  inertial instability}

\begin{figure}
\begin{center}
\includegraphics[width=8cm]{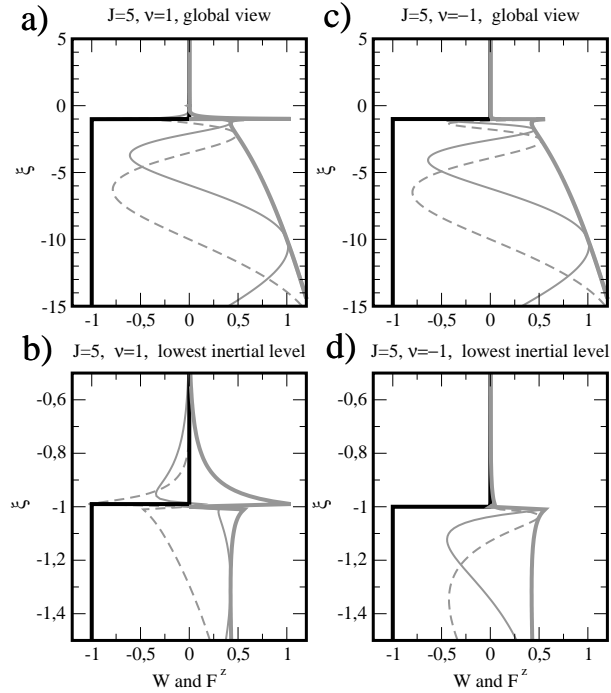}
\end{center}
\caption{Vertical velocity $W$ for the
exact solution~(\ref{eq:sol1})-(\ref{eq:sol3}) ($|W|$ thick grey solid, $\Re W$ grey solid, and $\Im W$ grey dashed) and  EP flux $F^z$
from (\ref{eq:EP_flux}) (black solid) for $J=5$ and $\nu=\pm 1$.
The EP flux is normalized by its incident value: (a) and (c) global views for $\nu=\pm 1$; (b) and (d) zooms on the lowest inertial level. }
\label{fig:EPflux_J5v0}
\end{figure}

To describe further the valve effect, we return to the exact solutions and first present vertical profiles of $W$ when $J=5$ and for $\nu=\pm1$ in Fig.~\ref{fig:EPflux_J5v0} . The global views in Figs.~\ref{fig:EPflux_J5v0}a and \ref{fig:EPflux_J5v0}c show that the solutions are almost identical. Even though a closer examination of the solution at the lowest inertial levels (Figs.~\ref{fig:EPflux_J5v0}b and \ref{fig:EPflux_J5v0}d) shows that the solution with $\nu>0$ is amplified whereas that with $\nu<0$ is attenuated, this difference matters little because of the very rapid decay with $\xi$ of both solutions between the inertial levels. Also in Fig.~\ref{fig:EPflux_J5v0}, we see that the EP flux is very small between the inertial levels consistent with  expression (\ref{eq:epratio}) for the ratio ${F^{z(\mathrm{II})}}/{|F^{z(\mathrm{I})}|}$ of the EP flux between the inertial level to the incident EP flux. For $J \gg 1$, this ratio is approximated by (\ref{eq:epratiowkb}) which shows, as discussed, that the valve effect plays only a minor role in modulating it.

\begin{figure}
\begin{center}
\includegraphics[width=8cm]{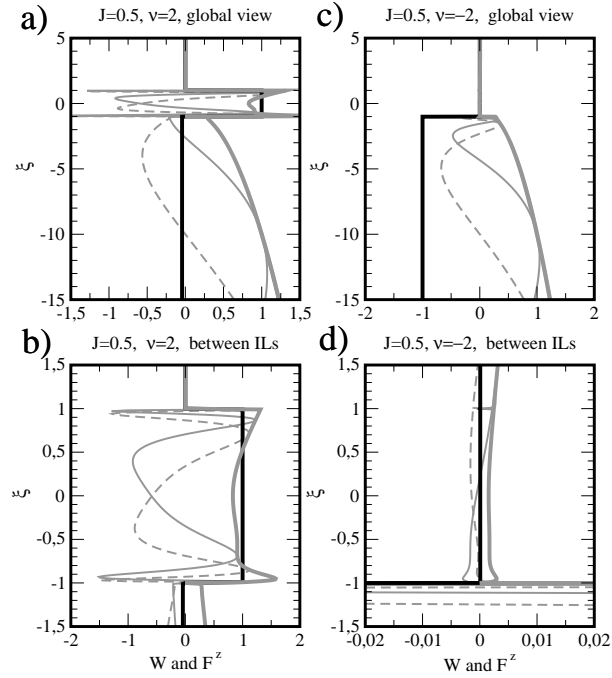}
\end{center}
\caption{Same as Fig.~\ref{fig:EPflux_J5v0} but for $J=0.5$, $\nu=\pm 2$. In a) and b), the EP flux is normalized by its amplitude between the inertial layers. Note the change of scale in panel (d).}
\label{fig:EPflux_J0v5}
\end{figure}

For $J<1$, the situation is different, and the valve effect can lead to an intensification of the perturbation between the inertial levels. This is illustrated in Figure~\ref{fig:EPflux_J0v5} which shows profiles of $W$ and $F^z$ for $J=0.5$ and $\nu=\pm 2$. When $\nu=2$ in Figs.~\ref{fig:EPflux_J0v5}a-b, the incident wave is clearly amplified as it passes through the lower inertial level (Fig.~\ref{fig:EPflux_J0v5}b), and the solution between the inertial levels does not display the almost exponential decay with altitude that characterizes the solution when $J\gg 1$. 
The behaviour for $\nu=-2$  illustrated in Figs.~\ref{fig:EPflux_J0v5}c-d is then completely different since the incident wave is almost entirely absorbed at the lowest inertial level. This pronounced  difference between positive and negative $\nu$ translates in the EP fluxes ratios in (\ref{eq:epratio})
as Fig.~\ref{fig:EP_FLUX} confirms. \red{
More importantly, (\ref{eq:epratio}) tells that the disturbance with $\nu>\sqrt{(J-1/4)/(1-J)}$ are amplified between the inertial levels when $J<1$ (in this case  the ratio is larger than $1$).}
 More than this, for \red{$|\nu| \gg 1$}, 
\red{
\begin{equation}
\frac{F^{z(\mathrm{II})}}{|F^{z(\mathrm{I})}|} \sim \e^{2\pi  (\nu-\sqrt{J}|\nu|)},
\end{equation}
so the amplification is arbitrarily large for $\nu \to +\infty$ when $J<1$. 
}

It is interesting to relate this result to  inertial
instability, which also occurs  when $J < 1$ (Stone 1966), a condition that implies that the background
flow potential vorticity is negative (Bennetts and Hoskins 1979).
Inertial instability -- or symmetric instability -- arises for perturbations
with $k=0$, that is, $\nu = \infty$, a case that
we have not treated explicitly.
It \red{is examined in Stone~(1966)} and can be recovered easily by replacing the non-dimensional variable
$\xi$ defined in (\ref{eq:xi}) and clearly inappropriate for $k=0$ by the original $z$.
This  reduces the equation for $W$ to  
\begin{equation} \label{eq:inertial}
\left(f^2 - \omega^2 \right) W_{zz} + 2 i \Lambda f l W_z - J \Lambda^2 l^2 W = 0  
\end{equation}
which admits the simple solution $W \propto \exp(i m z)$ for
\begin{equation}\label{eq:inertialb}
\omega^2 = f^2 \left(J \delta^2 + 2 \delta + 1 \right),
\end{equation}
where $\delta = (\Lambda l)/(f m)$. A purely imaginary frequency $\omega$, corresponding to symmetric instability, occurs for $J<1$, with maximum growth rate $\mathrm{Im} \, \omega = f \sqrt{1/J-1}$ for $\delta=-1/J$.

Although our 
analysis is not adapted to treat this case, because when $k \rightarrow 0$ the inertial levels go to infinity  and it becomes impossible to impose radiation conditions beyond them, the condition $J<1$ for IGW amplification between the inertial levels establishes a clear connection  between the valve effect and the 
inertial instability criteria. This connection with inertial instability is not the only one. For instance, the fact that
waves with large $\nu$ are amplified between the inertial levels
suggests that the shear layer response to a random incident IGW field
can become very large between the inertial levels and preferentially
symmetric as are inertial instabilities. \red{Also, the response in this transverse plane will be tilted in the direction of the isentropes (see Figs.~2d and 2h in LPV12) as is also the case for inertial instability} since $\delta<0$ for them (see (\ref{eq:inertialb})).
Note also that a stabilization of the
flow by the disturbance is expected to occur \red{in the nonlinear regime}. Indeed
the large increase of the EP flux at the lower inertial level
accelerates the mean flow there, whereas the decrease at the upper
inertial level  decelerates it there. As a result, the positive
shear  tends to be reduced: as is often the case, the development of the instability weakens its cause.

\section{Conclusion}

The problem of the absorptive properties of gravity waves by
critical levels analyzed by Booker and Bretherton~(1967)
is one of the cornerstone of dynamical meteorology.  
Its extension to the rotating case, even in the simple
set-up with constant shear, constant stratification
and without boundaries has been therefore studied by many investigators. 
Nevertheless, as the mathematics are involved,
its exact solution has never been
given adequately over the full domain.
This paper gives this solution and interprets it, partly using a WKB approximation.

In the general case, the structure of the solution is as follows.
A monochromatic wave incident at the lower inertial level
(there is no reflected wave) is amplified or attenuated 
by the valve effect, depending on the Richardson number and on the orientation of the wavevector. Between the inertial levels
the disturbance consists of two independent solutions (\ref{eq:sol2});
in the WKB as in the QG approximation, these two solutions are necessary to explain a non-zero EP flux. 
The amplification (attenuation) at the lower inertial level is compensated by an inverse attenuation (amplification) at the upper inertial level, so that the overall attenuation between the two inertial levels is controlled by the decrease of the solution across the region in a manner akin to quantum tunnelling. The analogy is not complete, however, in  particular because there is no reflected gravity wave.  

The situation is particularly transparent for large Richardson number $J \gg 1$, when a WKB approximation provides the solution in terms of elementary functions. In this case, the solution between the inertial levels is dominated by an exponentially decaying part, analogous to that predicted in the QG approximation. An exponentially small, exponentially growing solution is however present and ensures the constancy of the EP flux; this solution can be derived by careful consideration of the Stokes phenomenon (see LPV12 for details).

When $J$ is smaller, the decay of the balanced solution with
altitude is not as pronounced, and the valve effect becomes significant, leading to a strong dependence of the wave amplitude between the inertial levels on the wavevector orientation. For $J<1$, in particular, waves with large $\nu$,
experience an amplification, with an EP flux between the inertial levels that is larger than that of the incident wave.
This establishes a connection
between the absorptive properties of the inertial levels and the criterion $J<1$ for inertial (symmetric) instability. This connection is not apparent	either in the 
classical transmission coefficient in (\ref{eq:BB_Abs}) or in the formula
for the valve effect in (\ref{eq:JO_Abs}). Practically, this demonstrates 
how small non-symmetric disturbances can become very large in
inertially unstable flow, providing there is a small external
excitation.

\medskip

\noindent
\textbf{Acknowledgments.} 
This work was supported by the European Commission's 7th Framework Programme,
under the projects EMBRACE (Grant agreement 282672), and by the French ANR 
project StradyVarius. The authors acknowledge the International Space Science Institude in Bern for hosting two workshops about atmospheric gravity waves.
\red{We would also like to thank the three anonymous reviewers for their useful comments.}

\appendix
\section{\red{Derivation details}}
\subsection{Exact Derivation of $R$ and $T$}

Following LPV12, the $W$ functions in (\ref{eq:sol1})-(\ref{eq:sol3})
can be expressed in terms of hypergeometric functions $F$. 
\red{Using Eqs.\ (15.5.3), (15.5.4), (15.5.6) and (15.5.7) in 
Abramowitz and Stegun (1964) (hereafter AS)}, we write
\begin{subequations}
\begin{equation}
W_u^{(\mathrm{III})}=(1+\xi)^{-i\nu}\xi^{-2b}F(a',b';c';\xi^{-2}), \label{Eq1a_A1}
\end{equation}
\begin{equation}
W_d^{(\mathrm{II})}=(1+\xi)^{-i\nu}F(a,b;c;\xi^2), \;
W_u^{(\mathrm{II})}=(1+\xi)^{-i\nu}\xi^3\,F(a'',b'';c'';\xi^2),
\label{Eq1c_A1}
\end{equation}
\begin{equation}
W_d^{(\mathrm{I})}=\left(|\xi|-1\right)^{-i\nu}|\xi|^{-2a}F(a''', b''';c''';|\xi|^{-2}),
W_u^{(\mathrm{I})}=\left(|\xi|-1\right)^{-i\nu} |\xi|^{-2b}F(a', b';c';|\xi|^{-2})
\label{Eq1e_A1}
\end{equation}
\end{subequations}
In these expressions, the coefficients $a$, $b$ and $c$ are given by
\begin{equation}
a=-\frac{1}{4}-i\frac{\nu-\mu}{2},\quad
b=-\frac{1}{4}-i\frac{\nu+\mu}{2},\quad
c=-\frac{1}{2}, 
\label{Eq2_A1}
\end{equation}
and $\mu=\sqrt{J(1+\nu^2)-1/4}$.
The other coefficients are linear combinations of $a$, $b$ and $c$:
$a'=b$, $b'=b-c+1$, $c'=b-a+1$, $a''=a-c+1$, $b''=b-c+1$, $c''=2-c$,
$a'''=a$, $b'''=a-c+1$ and $c'''=a-b+1$.
Then to evaluate the complex constants $A$, $B$, $C$ and $D$ we \red{match (\ref{eq:sol1}) and (\ref{eq:sol2}) in the vicinity of $\xi=1$ and (\ref{eq:sol2}) and (\ref{eq:sol3}) in the vicinity of $\xi=-1$ along the contour displayed in Fig.~\ref{fig:TP} (see also Eq.~(\ref{eq:branchs})). Using (\ref{Eq1a_A1})-- (\ref{Eq1e_A1}) and
the transformation formula (15.3.6) in AS, we obtain}
\begin{subequations}
\begin{equation}
W\sim \alpha'(\xi-1)^{i\nu}+2^{-i\nu}\beta' \quad \textrm{as} \, \ \xi \to 1^+,
\label{Eq3a_A1}
\end{equation}
\begin{equation}
W\sim
\left(\alpha A+\alpha''B\right)(1-\xi)^{i\nu}
+2^{-i\nu}(\beta A+\beta'' B) \quad \textrm{as} \,  \ \xi \to 1^-,
\label{Eq3b_A1}
\end{equation}
\begin{equation}
W\sim 2^{i\nu}(A\alpha-B\alpha'') + \left(1+\xi\right)^{-i\nu}
\left(A\beta-B\beta''\right)\, \ \xi \to -1^+,
\label{Eq3c_A1}
\end{equation}
\begin{equation}
W\sim 2^{i\nu}(C\alpha'''+D\alpha') + \left(|\xi|-1\right)^{-i\nu}
\left(C\beta'''+D\beta'\right)
\, \ \xi \to -1^-.
\label{Eq3d_A1}
\end{equation}
\end{subequations}
In these expressions the $\alpha$'s and the $\beta$'s are related
to the $a, b, c$'s by
\begin{equation}
\alpha=\frac{\Gamma(c)\Gamma(a+b-c)}{\Gamma(a)\Gamma(b)}
\quad \textrm{and} \quad
\beta=\frac{\Gamma(c)\Gamma(c-a-b)}{\Gamma(c-a)\Gamma(c-b)},
\label{Eq4_A1}
\end{equation}
where $\Gamma$ is the gamma function~(see AS, chapter 6).
As a first step to \red{determine $A$, $B$, $C$ and $D$} it should be noticed
that the 8 coefficients $\alpha$'s and $\beta$'s can ultimately be expressed
in terms of the 4 coefficients $\alpha$, $\alpha'$, $\beta'$, and $\alpha''$, given explicitly as
\begin{subequations} \label{aaa}
\begin{equation}
\red{ 
\alpha'=\frac{\Gamma(1-i\mu)\Gamma(-i\nu)}{\Gamma(b)\Gamma(1-b^*)},\quad
\beta'=\frac{\Gamma(1-i\mu)\Gamma(i\nu)}{\Gamma(1-a)\Gamma(a^*)},}
\label{Eq5a_A1}
\end{equation}
\begin{equation}
\red{\alpha=\frac{-2 \sqrt{\pi}\Gamma(-i\nu)}{\Gamma(a)\Gamma(b)},\quad 
\alpha''=\frac{3\sqrt{\pi}\Gamma(-i\nu)}{4\Gamma(1-a^*)\Gamma(1-b^*)},}
\label{Eq5b_A1}
\end{equation}
\end{subequations}
through the relations $\beta=\alpha^*$, $\beta''=\alpha''^*$, $\alpha'''=\beta'^*$ and $\beta'''=\alpha'^*$.

Requiring that the continuations match~(\ref{Eq3b_A1}) and~(\ref{Eq3d_A1}) determines the constants $A$, $B$  and $C$, $D$ through
\red{Matching near $\xi=1$ and $\xi=-1$ yields the linear systems (\ref{eq:system1})--(\ref{eq:system2}).}
Solving~(\ref{eq:system1}) for $A$ and $B$ gives
\begin{subequations} \label{AAA}
\begin{equation}
\red{
A=\frac{\alpha''\beta'}{\delta_1}-\frac{\alpha''^*\alpha'}{\delta_1}\hbox{e}^{\pi\nu},
}
\label{Eq6a_A1}
\end{equation}
\begin{equation}
\red{ 
B=\frac{\alpha'\alpha^*}{\delta_1}\hbox{e}^{\pi\nu}-\frac{\alpha\beta'}{\delta_1}.
}
\label{Eq6b_A1}
\end{equation}
\end{subequations}
Solving~(\ref{eq:system2}) for $C$ and $D$ gives
\begin{subequations} \label{bbb}
\begin{equation}
\red{
C=\frac{\beta'(A\alpha-B\alpha'')}{\delta_2}-\frac{\alpha'(A\alpha^*-B\alpha''^*)}{\delta_2}\hbox{e}^{-\pi\nu},}
\label{Eq7a_A1}
\end{equation}
\begin{equation}
\red{
D=\frac{\beta'^*(A\alpha^*-B\alpha''^*)}{\delta_2}\hbox{e}^{-\pi\nu}-\frac{\alpha'^*(A\alpha-B\alpha'')}{\delta_2}.}
\label{Eq7b_A1}
\end{equation}
\end{subequations}
In these equations, $\delta_1$ and $\delta_2$ are  determinants appearing in (\ref{eq:system1})--(\ref{eq:system2}), explicitly given by
\begin{equation}
\red{\delta_1=\alpha^*\alpha''-\alpha\alpha''^*\quad \hbox{and}\quad
\delta_2=|\beta'|^2-|\alpha'|^2.}
\label{Eq8_A1}
\end{equation}
A closer examination of \red{(\ref{aaa})--(\ref{bbb})} shows that the Gamma functions \red{involved} appear through products expressible in terms of ordinary trigonometric functions
using the reflection formula
\begin{equation}
\Gamma(z)\Gamma(1-z)=\frac{\pi}{\sin \pi z}
\label{Eq9_A1}
\end{equation}
and the two related formulas (6.1.29) and (6.1.31) in AS.
\red{
\red{We now sketch the necessary calculation.}
For brevity we introduce the notation $s=\sin(\pi a^*)$ and $t=\sin(\pi b)$.
Applying~(\ref{Eq9_A1}) with $z=a^*$ and $z=b$ thus reduces the right-hand sides to $\pi s^{-1}$ and $\pi t^{-1}$, respectively.}
%

Substituting~(\ref{aaa}) into the two equations in~(\ref{Eq8_A1}) gives 
\begin{equation}
\red{
\delta_1=\frac{3}{2}\frac{ts^*-st^*}{\nu\sinh(\pi\nu)}\quad \hbox{and}\quad 
\delta_2=\frac{\mu}{\nu}\frac{|s|^2-|t|^2}{\sinh(\pi\nu)\sinh(\pi\mu)}.
}
\label{Eq10b_A1}
\end{equation}
These  equations can be further simplified using common trigonometric identities \red{and the definitions of $a$, $b$ and $c$ in terms
of $\nu$ and $\mu$.}
A direct calculation gives
\begin{equation}
\red{
ts^*-st^*=i\sinh(\pi \nu)\quad \hbox{and}\quad
|s|^2-|t|^2=-\sinh(\pi\mu)\sinh(\pi\nu),
}
\label{Eq11b_A1}
\end{equation}
reducing the \red{determinants} to 
\begin{equation}
\red{
\delta_1=\frac{3i}{2\nu}\quad \hbox{and}\quad \delta_2=-\frac{\mu}{\nu}.
}
\label{Eq0_rev}
\end{equation}

We can use this approach to obtain \red{exact} formulas for the transmission and reflection coefficients.
\red{
These coefficients are defined by $T=D^{-1}$ and $R=CD^{-1}$, which requires the calculation of $C$ and $D$.
Substituting~(\ref{Eq6a_A1}) and~(\ref{Eq6b_A1}) into~(\ref{Eq7b_A1}) gives
\begin{equation}
D=\frac{2\hbox{Re}(\alpha\alpha''^*)(|\beta'|^2\hbox{e}^{-\pi\nu}+|\alpha'|^2\hbox{e}^{\pi\nu})-
4\hbox{Re}(\alpha\alpha''\beta'\alpha'^*)}{\delta_1\delta_2},
\label{Eq1_rev}
\end{equation}
where the functions involving the constants $\alpha'$, $\beta'$, $\alpha$ and $\alpha''$ in the numerator 
can be expressed in terms of $s$ and $t$ as 
\begin{subequations} \label{ccc}
\begin{equation}
\alpha^*\alpha''=\delta_1\frac{st^*}{st^*-ts^*},
\label{Eq2a_rev}
\end{equation}
\begin{equation}
\alpha\alpha''\beta'\alpha'^*=-\delta_1\delta_2\frac{|s|^2|t|^2}{(|s|^2-|t|^2)(ts^*-st^*)},
\label{Eq2b_rev}
\end{equation}
\begin{equation}
|\alpha'|^2=\delta_2\frac{|t|^2}{|s|^2-|t|^2}\quad \hbox{and}\quad |\beta'|^2=\delta_2\frac{|s|^2}{|s|^2-|t|^2}. 
\label{Eq2c_rev}
\end{equation}
\end{subequations}
Substituting~(\ref{ccc}) into~(\ref{Eq1_rev}), we note that the factor $\delta_1\delta_2$ appears in both 
the numerator and the denominator. Using trigonometric identities, we finally obtain the simple form 
\begin{equation}
D=-i\e^{\pi\mu},
\label{Eq3_rev}
\end{equation}
or equivalently,  $T=D^{-1}=i\e^{-\pi\mu}$.
}

\red{
The derivation of a formula for the reflection coefficient follows the same argument.
An expression for $C$ is obtained by substituting~(\ref{AAA}) into~(\ref{Eq7a_A1}),
and the product of coefficients $\alpha'$, $\beta'$, $\alpha''$ is expressed in terms of the trigonometric functions $s$ and $t$,
 using the reflection formula.
Hence, we deduce \rev{
\begin{equation}
C=2\delta_1^{-1}\delta_2^{-1}\left[\alpha\alpha''\beta'^2+(\alpha\alpha'')^*\alpha'^2-
2\alpha'\beta'\cosh(\pi\nu)\hbox{Re}(\alpha^*\alpha'')\right].
\label{Eq4_rev}
\end{equation}
}
along with the two relations 
\begin{subequations}
\begin{equation}
\alpha\alpha''\beta'^2+\alpha'^2(\alpha\alpha'')^*=-\frac{2\pi}{3}\delta_1^2\frac{\Gamma^2(1-i\mu)}{P(a,b)}
\frac{|s|^2+|t|^2}{(ts^*-st^*)^2},
\end{equation}
\begin{equation}
2\alpha'\beta'\hbox{Re}(\alpha^*\alpha'')=-\frac{2\pi}{3}\delta_1^2\frac{\Gamma^2(1-i\mu)}{P(a,b)}
\frac{st^*+s^*t}{(ts^*-st^*)^2},
\end{equation}
\end{subequations}
where $P(a,b)=\Gamma(a^*)\Gamma(b)\Gamma(1-b^*)\Gamma(1-a)$.
After some direct algebra, we find
\begin{equation}
C=0,
\end{equation}
or equivalently, $R=0$,
which is valid to all orders in $\mu$.
}

%

\subsection{EP flux jump between the inertial levels}
\red{From (\ref{eq:sol2}), (\ref{Eq1c_A1}), and the definition of the Gauss 
hypergeometric series (15.1.1 in AS), we can show that near $\xi=0$
\begin{equation}
W= A\left(1-i\nu\xi-\left(\frac{\mu^2}{2}+\frac{1}{8}\right)\xi^2
-i\nu\left(\frac{\nu^2}{3}-\frac{\mu^2}{2}+\frac{5}{24}\right)\xi^3\right)
+B\xi^3+O(\xi^4).
\end{equation}
From  $F^z$ evaluated very near $\xi=0$ using (\ref{eq:EP_flux}), we deduce that $F^z=F^{z(\mathrm{II})}={3i}\left(BA^*-AB^*\right)/2$ between the inertial levels.
Similarly, using the asymptotics form for $W$ as $\xi \to -\infty$,
we deduce that 
$F^{z(I)}=-\mu DD^*$ below $\xi=-1$. The ratio
of the EP flux between the inertial levels to the incident EP flux is therefore given by}
\begin{equation}
\frac{F^{z(\mathrm{II})}}{|F^{z(\mathrm{I})}|}=\frac{3i}{2\mu}\frac{BA^*-AB^*}{DD^*}.
\label{Eq1_A2}
\end{equation}
According to \red{(\ref{Eq3_rev})}, the denominator of~(\ref{Eq1_A2}) can be simplified to $DD^*=\hbox{e}^{2\mu \pi}$, while
the numerator takes a more complicated form.
Using \red{(\ref{AAA})}, we obtain 
\begin{equation}
BA^*-AB^*=\frac{1}{|\delta_1|^2} \left(\hbox{e}^{2\pi\nu}|\alpha'|^2-|\beta'|^2\right)\left(\alpha\alpha''^*-\alpha^*\alpha''\right).
\label{Eq2_A2}
\end{equation}
%
%
%
Upon substituting~\red{(\ref{Eq2a_rev}) and~(\ref{Eq2c_rev})} into (\ref{Eq2_A2}), we obtain
\begin{equation}
BA^*-AB^*=-\frac{2\mu }{3}\frac{(s^*t-st^*)(\hbox{e}^{2\pi\nu}|t|^2-|s|^2)}{\sinh^2(\pi\nu)\sinh(\pi\mu)}.
\label{Eq4_A2}
\end{equation}
This equation can be further simplified using~(\ref{Eq11b_A1}) and the  identity
\begin{equation}
\hbox{e}^{2\pi\nu}|t|^2-|s|^2=\frac{1}{2}\hbox{e}^{\pi(\nu+\mu)}\sinh(2\pi\nu).
\label{Eq5_A2}
\end{equation}
Using this, (\ref{Eq1_A2}) reduces to
\begin{equation}
\frac{F^{z(\mathrm{II})}}{|F^{z(\mathrm{I})}|}=\frac{\cosh(\pi\nu)}{\sinh(\pi\mu)}\hbox{e}^{\pi(\nu-\mu)}=\frac{\hbox{e}^{2\pi\nu}+1}{\hbox{e}^{2\pi\mu}-1}.
\label{Eq6_A2}
\end{equation}
\red{
\subsection{\red{Continuation} of the WKB solution}
At the order $\mu$, substitution of the WKB solutions (\ref{eq:order1}) into  expression (\ref{eq:psidef}) for $W(\xi)$  when $\xi>1$ gives
\begin{equation}
W^{(\mathrm{III})}
(\xi)=\xi(\xi-1)^{-1/4+i\nu/2}
(\xi+1)^{-1/4-i\nu/2} \e^{i\mu\cosh^{-1}\xi},
\label{eq:WKB_details1}
\end{equation}
where $\cosh^{-1}\xi=\ln \left(\xi+\sqrt{\xi^2-1}\right)$. To continue this solution below $\xi=1$, we write  $(\xi-1)=(1-\xi) \e^{-i\pi}$ so 
$\cosh^{-1}\xi$ becomes $\ln \left(\xi-i\sqrt{1-\xi^2}\right)=i\sin^{-1}\xi-i{\pi}/{2}$. Therefore the continuation of $W^{(\mathrm{III})}$ reads
\rev{\begin{equation} 
 W^{(\mathrm{II})}
(\xi)=\xi(1-\xi)^{-1/4+i\nu/2}\e^{\nu\pi/2+i\pi/4}
(\xi+1)^{-1/4-i\nu/2} \e^{+\mu\pi/2-\mu\sin^{-1}\xi}.
\label{eq:WKB_details2}
\end{equation}
}
To continue this function below $\xi=-1$, we write  $\xi+1=|\xi+1|\e^{-i\pi}$ \rev{(and also $\xi=-|\xi|$)}, so
$\sin^{-1}\xi$ transforms into 
$-i\ln\left(|\xi|+\sqrt{\xi^2-1} \right)-\pi/2$,
leading to the continuation
 \begin{equation}
 W^{(\mathrm{I})}
(\xi)=-\xi(1-\xi)^{-1/4+i\nu/2}\e^{-i\pi/2}
|\xi+1|^{-1/4-i\nu/2} \e^{+\mu\pi+i\mu\ln \left(|\xi|+\sqrt{\xi^2-1}\right)}.
\label{eq:WKB_details3}
\end{equation}
The solutions in (\ref{eq:WKB_III})--(\ref{eq:WKB_I}) are just (\ref{eq:WKB_details1})--(\ref{eq:WKB_details3}) multiplied by $i \e^{-\mu\pi}$
so that the incident wave has a unit amplitude. 
}

\section*{References}
\begin{description}

\item 
Ablowitz M.J. and Fokas A.S., 1997
Complex variables: introduction and applications,
Cambridge University Press.

\item
Abramowitz M. and I.A. Stegun, 1964
Handbook of mathematical functions (9th edition),
{\em Dover Publications Inc., New York},
1045pp. 

\item
Alford, M.H., 2003, Redistribution of energy available for ocean mixing by long-range propagation of internal waves, {\em Nature,} {\bf 423,}
159-162. 

\item
Andrews, D.G., J.R. Holton and C. B. Leovy, 1987,
Middle atmosphere dynamics, {Academic Press Inc. (London) LTD,} 489p.

\item
Bender, C.M., and S.A. Orszag, 1978 Advanced mathematical methods for scientists and engineers, {\em McGraw-Hill int. book comp.}, 593pp.

\item
 Bennetts, D.A., and  B.J. Hoskins, 1979
Conditional symmetric instability -
a possible explanation for frontal rainbands,
{\em Quart. J. Roy. Met. Soc.,}
{\bf 105,}  945--962. 

\item
Booker, J.R. and F.P. Bretherton, 1967
The critical layer for internal gravity waves in a shear flow,
{\em J. Fluid Mech.}, {\bf 27}, 513--539.
%
%

\item
Eliassen, A. and E. Palm, 1961 On the transfer of energy in
stationary mountain waves, {\em Geofys. Publ.,}
{\bf 22,}
1--23.

\item
Ferrari, R., and C. Wunsch, 2009 Ocean circulation kinetic energy:
Reservoirs, sources and sinks, {\em Annu. Rev. Fluid. Mech,}
{\bf 41,} 253--282.

\item
Gill, A., 1982 Atmosphere-Ocean Dynamics, {\em Academic Press}, 662p

\item
Grimshaw, R, 1975 Internal gravity waves: critical layer absorption in a rotating fluid, {\em J. Fluid Mech.}, {\bf 70}, 287--304.

\item
Gula, J. and V. Zeitlin, 2010
Instabilities of buoyancy driven coastal currents and their nonlinear evolution in the two-layer rotating shallow water model. Part I: Passive lower layer,
{\em J. Fluid Mech.}, {\bf 659,} 69--93.

\item
Hertzog, A., G. Boccara, R.A. Vincent, F. Vial, and P. Cocquerez, 2008
 Estimation of gravity wave momentum flux and phase speeds
 from quasi-Lagrangian stratospheric balloon flights.
Part II: Results from the Vorcore campaign in Antarctica,
{\em J. Atmos. Sci.}, {\bf 65,} 3056--3070.

\item
Howard, L.H., 1961 Note on a paper of John W. Miles,  {\em J. Fluid. Mech,}
{\bf 10,} 509-512.

\item
Inverarity, G.W. and G.J. Shutts, 2000
A general, linearized vertical structure equation for the vertical
velocity: Properties, scalings and special cases,
{\em Quart. J. Roy. Meteor. Soc.}, {\bf 126}, 2709--2724.

\item
Jones, W.L., 1967
 Propagation of internal gravity waves in fluids
 with shear flow and rotation, {\em J. Fluid Mech.}{\bf , 30,} 439--448.

\item
Lindzen, R.S., 1988 Instability of plane parallel shear-flow (toward a mechanistic picture of how it works), {\em Pure Applied Geophys.}, {\bf 126,} 103--121.

\item
Lott, F., H. Kelder, and H. Teitelbaum, 1992
A transition from Kelvin-Helmholtz instabilities to propagating wave
  instabilities,
  {\em Phys. Fluids,}
  {\bf 4,} 1990--1997.

\item
Lott, F., 1997
 The transient emission of propagating gravity waves by a stably
 stratified shear layer,
{\em Quart. J. Roy. Meteor. Soc.,}
{\bf 123}, 1603--1619.

\item
Lott, F., 2003
Large-scale flow response to short gravity waves breaking in a rotating
  shear flow,
{\em J. Atmos. Sci.,} {\bf 60,} 1691--1704.
\item

Lott F., R. Plougonven, and J. Vanneste, 2010
Gravity waves generated by sheared potential-vorticity anomalies,
{\em J. Atmos. Sci.,} {\bf 67,} 157--170.

\item
Lott, F., R. Plougonven, and J. Vanneste, 2012:
Gravity waves generated by sheared three dimensional potential vorticity
anomalies, {\em J. Atmos. Sci.,} {\bf  69,} 2134–-2151

\item
Mamatsashvili, G.R., V.S. Avsarkisov, G.D. Chagelishvili, R.G. Chanishvili,and M. V. Kalashnik, 2010
Transient dynamics of nonsymmetric perturbations
versus symmetric instability in baroclinic zonal shear flows, 
{\em J. Atmos. Sci.,} {\bf 67,} 2972--2989.

\item
Marshall, J., and Coauthors, 2009 The CLIMODE field campaign:
Observing the cycle of convection and restratification over the Gulf Stream.
{\em Bull. Amer. Meteor. Soc.}, {\bf 90}, 1337--1350.

\item
McWilliams J.C., 2003
Diagnostic force balance and its limits, 
{\em  in Nonlinear Processes in Geophysical Fluid Dynamics,
 Kluwer, ed:  O. U. Velasco Fuentes, J. Sheinbaum and J. Ochoa,}
287--304.

\item
Miles, J.W., 1961: On the stability of heterogeneous shear flows, {\em J. Fluid. Mech,} {\bf 10,} 496--508.

\item
Miyahara, S., 1981
A note on the behavior of waves around the inertio frequency,
{\em J. Met. Soc. Japan,} {\bf 59,} 902--905.

\item
Molemaker, M.J., J. C. McWilliams, and I. Yavneh, 2005,
Baroclinic instability and loss of balance,
{\em J. Phys. Ocean.}, {\bf 35,} 1505--1517.

\item
Plougonven, R., and C. Snyder, 2007 Inertia-gravity waves spontaneously
generated by jets and fronts. Part I: Different baroclinic life cycles,
{\em J. Atmos. Sci.,} {\bf 64,} 2502--2520.

\item
Plougonven, R., D.J. Muraki,  and C. Snyder, 2005
A baroclinic instability that couples balanced motions and gravity waves,
{\em J. Atmos. Sci.,} {\bf 62,} 1545--1559.  

\item
Rabinovitch, A., O.M. Umurhan, N. Harnik, F. Lott, and E. Heifetz 2011 Vorticity inversion and action-at-a-distance instability in stably stratified shear flow, {\em J. Fluid Mech.,} {\bf 670,}  301--325.

\item
Richter, J.H., F. Sassi, and R.R. Garcia, 2010
Toward a physically based gravity wave source parameterization
in a General Circulation Model, 
{\em J. Atmos. Sci.,} {\bf 67,} 136--156.

\item
Sakai, S., 1989 Rossby--Kelvin instability: a new type of ageostrophic instability caused by a resonnance between Rossby waves and gravity waves,
{\em J. Fluid Mech.,} {\bf 202,} 149--176.

\item
Sato, K. and M. Yoshiki, 2008 Gravity wave generation around the
polar vortex in the stratosphere revealed by 3-hourly
radiosonde observations at Syowa station, 
{\em J. Atmos. Sci.,} {\bf 65}, 3719--3735.

\item
Scavuzzo, C.M., M.A. Lamfri, H. Teitelbaum and F. Lott, 1998
 A study of the
 low frequency inertio-gravity waves observed during {PYREX},
{\em J. Geophys. Res.}, {\bf D2}, {\bf 103}, 1747--1758.

\item
Shen, B.W., and Y.L. Lin, 1999
Effects of critical levels on two-dimensional back-sheared flow over
an isolated mountain ridge on an f plane,
{\em J. Atmos. Sci.,} {\bf 56,} 3286--3302.

\item
Shutts G.J., 2003 Inertia--gravity wave and neutral Eady wave trains
forced by directionally sheared flow over isolated hills,
{\em J. Atmos. Sci.}, {\bf 60,} 593--606.

\item
Stone, P.H., 1966
On non-geostrophic baroclinic instability, {\em J. Atmos. Sci.}, {\bf  23,}
390--400.

\item
Sutyrin G.G., 2008 Lack of balance in continuously stratified rotating flows,
{\em J. Fluid Mech.,} {\bf 615,} 93--100.

\item
Vanneste J. and I. Yavneh, 2007 
Unbalanced instabilities of rapidly rotating stratified sheared flows,
{\em J. Fluid Mech.,} {\bf 584,} 373--396.

\item
 Whitt, D.B. and L.N. Thomas, 2013
Near-inertial waves in strongly baroclinic currents.
 {\em J. Phys. Oceanogr.}, {\bf  43}, 706--725.

\item
Yamanaka, M. D., and H. Tanaka, 1984
Propagation and breakdown of internal inertia-gravity waves near critical levels in the middle atmosphere,
{\em J. Met. Soc. Japan}, {\bf 62,} 1--17.

\item
Zuelicke, C. and D. Peters, 2008
Parameterization of strong stratospheric inertia-gravity waves forced
by poleward-breaking Rossby waves,
{\em Month. Weath. Rev.,} {\bf 136,} 98--119.

\end{description}

\end{document}